\documentclass[%
 twocolumn,
 amsmath, amssymb,
 aps,
 pre,
 longbibliography
]{revtex4-1}

\usepackage{graphicx}
\usepackage{dcolumn}
\usepackage{bm}
\usepackage[english]{babel}
\usepackage[pdfpagelabels,plainpages=false,bookmarks=true,colorlinks,linkcolor=red,urlcolor=blue,citecolor=blue]{hyperref}

\usepackage{booktabs,subcaption,amsfonts,dcolumn}
\usepackage{multirow}
\usepackage{diagbox}
\renewcommand{\vec}[1]{{\bf #1}}

\newcommand{\braket}[1]{\langle #1  \rangle}
\newcommand{\bra}[1]{\langle #1 |}
\newcommand{\ket}[1]{| #1  \rangle}

\renewcommand{\max}{\text{max}}

\newcommand{\nn}{\text{nn}}
\newcommand{\nnn}{\text{nnn}}

\newcommand{\Ising}{\text{Ising}}

\newcommand{\abs}[1]{|#1|}

\newcommand{\Tr}{\text{Tr}}

\renewcommand{\v}{\vec{v}}

\begin{document}

\title{Continuous-time Monte Carlo Renormalization Group}

\author{Yantao Wu$^1$ and Roberto Car$^{1, 2}$}

\affiliation{
$^1$The Department of Physics, Princeton University\\
$^2$The Department of Chemistry, Princeton University \\
}

\date{\today}
\begin{abstract}
We implement Monte Carlo Renormalization Group (MCRG) in the continuous-time Monte Carlo simulation of a quantum system. 
We demonstrate numerically the emergent isotropy between space and time at large distances for the systems that exhibit Lorentz invariance at quantum criticality. 
This allows us to estimate accurately the sound velocity for these quantum systems.  
$Q$-state Potts models in one and two space dimensions are used to illustrate the method. 
\end{abstract}

\pacs{Valid PACS appear here}
\maketitle
\section{Introduction}

In recent years, continuous-time quantum Monte Carlo (QMC) has become the standard to simulate sign-free lattice quantum spin systems \cite{Continuous_time_I, Continuous_time_II, Diag_MC, Diag_MC_SciPost}.  
In continuous-time QMC, one adopts a path-integral representation, in which the time dimension is represented by a genuine continuous line of length equal to the inverse temperature $\beta$, whereas the space dimension is represented by a discrete lattice.
For systems that exhibit Lorentz invariance at criticality, the path integral represents a statistical field theory, in which isotropy between the time and space directions emerges at large distance. 
This occurs in systems whose low-lying elementary energy excitations depend linearly on the magnitude of the momentum of the excitation:  
\begin{equation}
  E(\vec k) - E_0 = v_s \abs{\vec k}.   
  \label{eq:dispersion}
\end{equation}
Here, $E_0$ is the ground state energy, and $\vec k$ is the momentum of a low-lying energy eigenstate of energy $E(\vec k)$.   
The constant of proportionality, $v_s$, is called the sound velocity and is a non-universal constant specific to a system.    
The sound velocity is the scale factor connecting time and length scales of a system.  
When Eq. \ref{eq:dispersion} happens, the large-distance behavior of a lattice spin system is described by a massless quantum field theory, invariant under dilational coordinate transformations. 
The invariance under dilational transformation underlies the success of renormalization group (RG) theory \cite{Wilson_RG} in treating critical phenomena, as done in the real space Monte Carlo Renormalization Group (MCRG) approach \cite{MCRG_SW}, which is often used to estimate critical couplings and exponents of spin systems.
When implemented on a discrete lattice, the dilational transformation is limited to integral scale factors.  
However, non-integral scale factors are desirable in many cases. 
For example, anisotropic coarse graining can lead to a fixed point Hamiltonian isotropic in the space and time directions. Then, the ratio of space and time lengths in the anisotropic coarse graining is the sound velocity.
The value of the sound velocity is generally a real number, requiring that the coarse-graining either along the space or the time direction be continuous.   
We show that this can be realized, within the MCRG framework, for quantum spin models on a lattice by adopting a continuous imaginary time path integral representation. In this way, Lorentz invariance is demonstrated numerically and accurate estimates of the sound velocity are obtained. In addition, continuous time MCRG allows us to compute the lattice version of the energy stress tensor of the underlying field theory. 

Conformal invariance is very powerful in two-dimensions (2D), due to the infinite dimensionality of the local conformal algebra \cite{BPZ}.  
Conformal field theory (CFT) yields finite sizing scaling predictions of physical observables, such as the energy \cite{Affleck_CFT, Blote_CFT} and the entanglement entropy \cite{EE_QFT}.      
Sound velocity and the energy-stress tensor are often parameters in these predictions. 
Thus, assuming the validity of these predictions, the sound velocity and the energy-stress tensor may be obtained by fitting observables in a numerical simulation against the CFT predictions. 
This has been carried out in many studies, e.g. \cite{Gehlen_CFT, Ma_CFT}.
With continuous-time MCRG, we determine these quantities from their defining expressions, without recourse to CFT results. 
This allows a much easier generalization to three dimensions (3D), where CFT predictions are less available. 
In this paper, we illustrate the idea of continuous-time MCRG mostly with examples in 2D, showing that already in 2D our approach leads to estimates of the sound velocity that are more accurate than those obtained with alternative methods, such as directly computing the energy spectrum or fitting numerical observables against CFT predictions.   
We also provide a 3D example, by reporting a calculation of the sound velocity for the quantum Ising model in two space dimensions, a system for which, to the best of our knowledge, the sound velocity cannot be obtained with other means.
A more detailed study of systems in 3D is deferred to future works. 

The paper is organized as follows. 
In Sec. \ref{sec:QMC}, we use a diagrammatic expansion to obtain the path integral representation of the partition function of the $Q$-state Potts model, the system that we use here as an example to illustrate the methodology.
In Sec. \ref{sec:mcrg}, we coarse grain the time direction and explain the MCRG procedure.   
In Sec. \ref{sec:vs}, we use continuous-time MCRG to compute the sound velocity.  
In Sec. \ref{sec:stress}, we interpret the coarse-graining along the time direction as a continuous coordinate transformation and discuss its connection with the energy-stress tensor. 
In Sec. \ref{sec:conclude}, we report our conclusions. 

\section{The Continuous-time Monte Carlo}
For concreteness, we illustrate the method with the $Q$-state Potts model having Hamiltonian \cite{Solyom_Potts}:
\begin{equation}
  \hat{H}_{Q} = - \sum_{\braket{i, j}} \sum_{k=0}^{Q-1} \hat\Omega_i^k \hat\Omega_{j}^{Q-k} - g \sum_{i=1}^{L^d} \sum_{k=1}^{Q-1} \hat M_i^k, 
\label{sec:QMC}
\end{equation}
where the system is in a $d$-dimensional hypercubic lattice with length $L$. 
$i$ and $j$ label different lattice sites, and $\braket{i,j}$ denotes a nearest neighbor bond. 
The operators $\hat{\Omega}_i$ and $\hat{M}_i$ act on the $Q$ states of the local Hilbert space at site $i$, which we label by $\ket{0}_i,...,\ket{s}_i,...\ket{Q-1}_i$.  
In this local basis, the $\hat\Omega_i$ is a diagonal matrix such that $\hat\Omega_i \ket{s}_i = \omega^{s} \ket{s}_i$, where $\omega = e^{i2\pi/Q}$ and $s = 0, \cdots, Q-1$. 
$\hat{M}_i$ performs a cyclic permutation: $\ket{0}_i\rightarrow \ket{Q-1}_i, \ket{1}_i \rightarrow \ket{0}_i, \cdots, \ket{Q-1}_i \rightarrow \ket{Q-2}_i$, and acts as a transverse-field.   
When $d=1$, this model is self-dual and a quantum phase transition occurs at the critical coupling $g_c = 1$ for all $Q$ \cite{Solyom_Potts}.  
When $Q \le 4$, the transition is continuous and is described by a CFT \cite{Baxter_Potts, Gehlen_CFT}. 
When $Q > 4$, the transition is first-order and has a finite latent heat at $g_c$ \cite{Baxter_Potts}.   

To derive a path-integral representation of the system partition function $Z = \Tr(e^{-\beta \hat H_Q})$, one takes 
\begin{equation}
  \hat H_0 = - \sum_{\braket{i, j}} \sum_{k=0}^{Q-1} \hat\Omega_i^k \hat\Omega_{j}^{Q-k}, \hspace{5mm} \hat H_1 = -g \sum_{i=1}^{L^d} \sum_{k=1}^{Q-1} \hat M_i^k 
\end{equation}
and considers the partition function in the interaction picture 
\begin{equation}
  Z = \Tr[\exp(-\beta \hat H_0) \hat T\{\exp(-\int_0^\beta \hat H_1(\tau) d\tau)\}],
  \label{eq:Z}
\end{equation}
where $\hat T$ is the time-ordering operator in imaginary time $\tau$, and $\hat H_1(\tau) = e^{\tau \hat H_0} \hat H_1 e^{-\tau \hat H_0}$ is the $\hat H_1$, in the interaction picture. 
Eq. \ref{eq:Z} can be written as a diagrammatic expansion in the following way:
\begin{widetext}
\begin{equation}
  \begin{split}
  Z &= \sum_{\bm\sigma} \sum_{n=0}^\infty \frac{(-1)^n}{n!} \bra{\bm\sigma} e^{-\beta \hat H_0}\hat T 
  \int_0^\beta \hat H_1(\tau_n)d\tau_n\cdots\int_0^\beta \hat H_1(\tau_1)d\tau_1\ket{\bm\sigma}
  \\
  &= \sum_{\bm\sigma} \sum_{n=0}^\infty g^n \int_0^\beta d\tau_n\int_0^{\tau_{n}} d\tau_{n-1} \cdots \int_0^{\tau_{2}}d\tau_1 \bra{\bm\sigma} e^{-(\beta-\tau_n)\hat H_0} \sum_{i=1}^{L^d} \sum_{k=1}^{Q-1} \hat M_i^k e^{-(\tau_n-\tau_{n-1})\hat H_0} \cdots e^{-\tau_1\hat H_0} \ket{\bm\sigma}
  \\
  &= \sum_{\bm\sigma} \sum_{n=0}^\infty g^n \sum_{i_1\cdots i_n}\int_0^\beta d\tau_n\int_0^{\tau_n} d\tau_{n-1} \cdots \int_0^{\tau_{2}}d\tau_1 \bra{\bm\sigma} e^{-(\beta-\tau_n)\hat H_0} \sum_{k=1}^{Q-1} \hat M_{i_n}^k e^{-(\tau_n -\tau_{n-1})\hat H_0} \cdots \sum_{k=1}^{Q-1} \hat M_{i_1}^ke^{-\tau_1\hat H_0} \ket{\bm\sigma}
\end{split}
\label{eq:diag}
\end{equation}
\end{widetext}
Here the states $\ket{\bm\sigma} = \otimes_i \ket{\sigma}_i$ form a basis in the Hilbert space.  
Each index $i_1,\cdots, i_n$ runs from $1$ to $L^d$. 
$\hat H_0$ is diagonal in the $\ket{\bm\sigma}$ basis: $\hat H_0\ket{\bm\sigma} = Q \sum_{\braket{i,j}} \delta_{\sigma_i,\sigma_j}\ket{\bm\sigma} \equiv h_0(\bm\sigma)\ket{\bm\sigma}$.  
Eq. \ref{eq:diag} suggests the following Monte Carlo (MC) scheme to sample the partition function.
For each $i = 1, \cdots, L^d$ and each $\tau \in [0,\beta]$, a Potts spin, $\sigma_i(\tau)$, ranging from $0$ to $Q-1$, is assigned to an MC configuration.   
As the state $\ket{\bm\sigma}$ is propagated in imaginary time, spin flips can happen at any lattice site and at any time. 
Let $\tau_l$ and $i_l$ denote the $l$th flip time and its associated lattice site. 
Here $l$ could be equal to $1, 2, \cdots,$ or $n$. 
In addition, let $\tau_l^-$ and $\tau_l^+$ respectively denote the time immediately before and after the flip time $\tau_l$. 
If a spin flip occurs at $\tau_l$ on site $i_l$, $\sigma_{i_l}(\tau_l^-)$ will be made to switch to any $\sigma_{i_l}(\tau_l^+)$ different from $\sigma_{i_l}(\tau_l^-)$ by the action of $\sum_{k=1}^{Q-1}\hat M_{i_l}^k$.  
In Eq. \ref{eq:diag}, the earliest spin flip occurs at $\tau_1$ on site $i_1$; the second one occurs at $\tau_2$ on site $i_2$, etc.. 
The total number of spin flips, $n$, contributes a weight $g^n$ to the sampling of the diagrammatic expansion.    
In addition, the weight includes factors equal to $e^{-(\tau_{l+1}-\tau_{l})h_0(\bm\sigma(\tau_l^+))}$ between two consecutive spin flips at $\tau_l$ ad $\tau_{l+1}$. 
Finally, the periodicity of the trace requires $\bm\sigma(\beta) = \bm\sigma (0)$, which in turn implies that $n$ should be even.
Thus, MC sampling does not have a sign problem even if $g$ is negative. 

The partition function in Eq. \ref{eq:diag} is given by a sum of terms (diagrams) that entail summation over discrete variables and integration over continuous ones. 
The contribution of the different terms, which are associated to the weights detailed above, is evaluated stochastically with a MC algorithm that follows the protocols discussed in Refs. \cite{Continuous_time_I, Continuous_time_II, Diag_MC}. 
For the $Q$-state Potts model diagrammatic MC can use a continuous time cluster algorithm \cite{Continuous_time_II}, based on the Wolff algorithm \cite{Wolff}, which significantly reduces equilibration time. 
We will use both local and cluster MC algorithms in the following. 

\section{The MCRG procedure}
\label{sec:mcrg}
Eq. \ref{eq:diag} indicates that the thermodynamics of a $d$-dimensional quantum Potts model is described by a statistical field theory in $(d+1)$ dimensions, where on each lattice site $i$ there is a {\it worldline} of length $\beta$ described by the function $\sigma_i(\tau)$.  
We can coarse grain this worldline by placing $P$ lattice points along the time direction through the majority-rule.
That is, we partition the worldline into $P$ intervals: $[0, \frac{P}{\beta}], [\frac{P}{\beta}, \frac{2\beta}{P}], \cdots, [\frac{(P-1)\beta}{P}, \beta]$.    
In each MC configuration, to each interval, we assign the Potts spin which appears most often on that interval.  
By discretizing time in this way we represent each worldline with $P$ discrete Potts spins and we end up with a ($d$+1)-dimensional hypercubic lattice that hosts $PL^d$ Potts spin. Each MC snapshot corresponds to a configuration of those spins.
The probability distribution of the spin configurations on the discrete lattice is not known explicitly, but can be sampled by coarse-graining the configurations generated in the diagrammatic MC simulation \cite{Continuous_time_II}.       
We can then perform MCRG on the $(d+1)$-dimensional hypercubic lattice. 
We designate the $n$-th RG iteration with a subscript $(n)$, where $n=0,1,2,..$. 
In the 0-th iteration the spin configuration, $\bm\sigma^{(0)}$, is the one obtained from coarse graining the time direction of diagrammatic MC. 
The probability distribution of $\bm\sigma^{(0)}$ is described by the (unknown) lattice Hamiltonian $H^{(0)}$. 
The subsequent levels of coarse graining are generated by successive block spin RG transformations and are characterized by spin configurations $\bm\sigma^{(n)}$ and Hamiltonians $H^{(n)}$. 

In all of the above coarse-graining transformations we use short-ranged coupling terms $S_\alpha(\bm\sigma)$ to parametrize the probability distribution $P(\bm\sigma)$ of the spin configurations $\bm\sigma$:  
\begin{equation}
  P(\bm \sigma) \propto e^{-H(\bm\sigma)}, \text{ where } H(\bm\sigma) = -\sum_{\alpha} K_\alpha S_\alpha(\bm\sigma), 
  \label{eq:p_sigma}
\end{equation}
with coupling constants $K_\alpha$. 
The terms $S_\alpha (\bm\sigma)$ include nearest neighbor ($\nn$), next nearest neighbor ($\nnn$), smallest plaquette ($\square$), etc., interactions. 

On the hypercubic lattice, a dilation transformation with scale factor $b$ is realized through a block spin transformation by a conditional probability, $T(\bm\mu|\bm\sigma)$, of the renormalized spin $\bm\mu$ given unrenormalized spins $\bm\sigma$.    
For example, in the majority-rule coarse-graining that we use, $\bm\sigma$ is partitioned into hypercubic blocks with side $b$. 
When there is a unique spin $\sigma_\max$ that appears the most often in the block, the renormalized spin $\mu$ for that block is assigned to $\sigma_\max$ with probability one. 
When $p$ $\sigma$-spins tie for the most appearances in the block, $\mu$ is assigned to one of these $p$ spins with probability $\frac{1}{p}$.  
Assuming the lattice is large enough, the spin configurations may be iteratively coarse-grained $n$ times, from $\bm\sigma^{(0)}$ to $\bm\sigma^{(n)}$, giving rise to a renormalized Hamiltionian at the $n$th RG level:   
\begin{equation}
  \label{eq:rg}
  \begin{split}
    H^{(n)}(\bm\sigma^{(n)}) &\equiv -\ln \sum_{\bm\sigma^{(0)}} T^{(n)}(\bm\sigma^{(n)}| \bm\sigma^{(0)}) e^{-H^{(0)}(\bm\sigma^{(0)})}  
\\ 
&= -\sum_{\alpha} K^{(n)}_\alpha S_\alpha(\bm \sigma^{(n)}) 
\end{split}
\end{equation}
where 
$T^{(n)}(\bm\sigma^{(n)}|\bm\sigma^{(0)})$ is the conditional probability of $\bm\sigma^{(n)}$ given $\bm\sigma^{(0)}$.  
$T^{(n)}(\bm\sigma^{(n)}|\bm\sigma^{(0)})$ implements coarse-graining from $\bm\sigma^{(0)}$ to $\bm\sigma^{(n)}$, realizing a dilation transformation with scale factor $b^n$.
It is obtained by iterating the single-level coarse-graining $n$ times:   
\begin{equation}
  \begin{split}
  T^{(n)}&(\bm\sigma^{(n)}|\bm\sigma^{(0)}) 
  \\
  &= \sum_{\bm\sigma^{(n-1)}}..\sum_{\bm\sigma^{(1)}} T(\bm\sigma^{(n)}|\bm\sigma^{(n-1)}) \cdots T(\bm\sigma^{(1)}|\bm\sigma^{(0)})
\end{split}
\end{equation}

In principle, an infinite number of coupling terms $S_\alpha(\bm\sigma^{(n)})$ is required to exactly parametrize $H^{(n)}$. 
In practice, this is not possible, and we only use a few coupling terms, such as those listed in Table \ref{table:tfim_1d_K}. 
Given a set of coupling terms, we use Variational Monte Carlo Renormalization Group (VMCRG) \cite{VMCRG} to compute the corresponding coupling constants $K^{(n)}_\alpha$. 
VMCRG allows us to compute the renormalized coupling constants in a way that greatly alleviates the critical slowing down and has very little finite-size effect \cite{CMTS}.
Its implementation details are explained in \cite{VMCRG, CMTS, TN_CMTS}.  
Here by finite-size effect, we mean the effect of different system size, $L$, with the same RG level, $n$. 
Our result, however, necessarily depends on what $n$ one uses to approach the fixed-point Hamiltonian. 
Using only a finite number of coupling terms in the Hamiltonian in Eq. \ref{eq:p_sigma} introduces a truncation errors. 
However, because the truncation scheme respects isotropy, i.e. the truncation of an isotropic Hamiltonian is still isotropic, we can estimate the sound velocity without truncation errors. 

\section{The sound velocity}
\label{sec:vs}
The sound velocity $v_s$ is an important property of a quantum system whose low lying excited states show linear momentum dispersion.  
In particular, this quantity is required to compare the predictions from CFT with observables of a lattice model.  
When $d = 1$, one can compute the sound velocity with finite size scaling of the ground state energy or entanglement entropy, assuming validity of the CFT prediction \cite{Affleck_CFT, Blote_CFT, EE_QFT}, or one can directly compute the excitation spectrum of the system. 
The latter calculation can be done by exact diagonalization of the system Hamiltonian, which is limited to small lattice sizes, or it can be done with density matrix renormalization group (DMRG) techniques \cite{DMRG}, which introduce truncation errors due to finite bond dimension. 
When $d > 1$, neither method works reliably, and one has to resort to QMC.   
In fact, the sound velocity has been calculated with continuous-time QMC by looking for a scale factor $v_s$ such that the coorelation function $C(x, \v_s \tau)$ becomes isotropic along the time and space directions at large distances \cite{Worm,CFT_3DIsing}.  
In the following, we will use directly RG to compute the sound velocity and show that VMCRG can deal with rather large lattice sizes, up to at least $L=256$, leading to very accurate estimates of the sound velocity.

To compute the sound velocity, we perform a dilation transformation with $b = 2$ using the majority rule.
In the VMCRG calculation, we take couplings along space and time directions to be independent, as the system is necessarily anisotropic between space and time.  
However, in the presence of Lorentz invariance, isotropy between space and time is recovered at large distances up to a scale factor $v_s$.    
One can thus vary $\frac{P}{\beta}$ until the couplings along the space direction and those along the time direction become equal for a large $n$. 
The $\frac{P}{\beta}$ so determined is the sound velocity, $v_s$.

\subsection{$Q=2$: The Ising model}
When $Q=2$, $\hat H_{Q=2}$ coincides up to an additive constant with the Hamiltonian of the transverse-field Ising model (TFIM), 
\begin{equation}
  \hat H_\Ising = -\sum_{\braket{i,j}} \hat\sigma_i^z\hat\sigma_j^z - g \sum_{i = 1}^{L^d} \hat\sigma_i^x, 
\end{equation}
where $\hat\sigma_i^{z,x}$ are the Pauli matrices at site $i$. 
We carry out the discussion using the terminology of the Ising model, i.e. instead of Potts spin $\sigma = 0, 1$, we will speak of Ising spin $\sigma = -1, 1$.    

The sound velocity for the one-dimensional TFIM is known exactly from its fermionic solution, and is $v_s = 2$ \cite{Kogut} :
\begin{equation}
  E(k) = 2 \sqrt{1 - 2 g_c \cos k + g_c^2} = 2 \abs{k} + o(\abs{k}), 
\end{equation}
where $g_c = 1$. 
Note that sometimes the Ising Hamiltonian is written with spin operators $\hat{S} = \frac{1}{2}\hat{\sigma}$, in which case $v_s$ would be $\frac{1}{2}$.
The $K^{(n)}$s calculated with VMCRG for the $d=1$ TFIM, simulated at $g_c$, are presented in Table \ref{table:tfim_1d}, for the coupling terms listed in Table \ref{table:tfim_1d_K}.  
We present the data for all the RG iterations from $n = 0$ to $5$ to illustrate the method. 
As clearly seen, the convergence to isotropy occurs with increasing $n$ when $\frac{P}{\beta} = v_s$.  
Note that the relative magnitude of $K^{(n)}_{1,x}$ and $K^{(n)}_{1,y}$ switches when $\frac{P}{\beta}$ crosses $2.0$.  
This gives an estimation of $v_s$ in the interval $(1.997, 2.003)$, to be compared, for example, with the DMRG result of $2.04$ found in \cite{Chepiga_Spectrum}. 
 
\begin{table}
  \setlength{\tabcolsep}{0.2em}
\centering
  \begin{tabular}{ll} 
    \hline
    \hline
    $\alpha$\hspace{5mm} & Coupling term\\
    \hline
    1$,x$  & nearest neighbor product along the time direction \\ 
    1$,y$  & nearest neighbor product along the space direction \\ 
    2      & 2nd neighbor product\\ 
    3      & product of spins in the smallest plaquette\\ 
    4$,x$  & 3rd neighbor product along the time direction\\ 
    4$,y$  & 3rd neighbor product along the space direction\\ 
    \hline
    \hline
  \end{tabular}
  \caption{The couplings used for $d = 1$ TFIM.
  Note that when $\alpha = 2$ and $3$, the couplings are themselves isotropic between space and time. 
  }
\label{table:tfim_1d_K}
\end{table}
\begin{table}
  \setlength{\tabcolsep}{0.8em}
\centering
\begin{subtable}[t]{0.48\textwidth}
  \begin{tabular}{lllll} 
    \hline
    \hline
    $n$ & $K^{(n)}_{1,x}$ & $K^{(n)}_{1,y}$ & $K^{(n)}_{4,x}$ & $K^{(n)}_{4,y}$\\
    \hline
    0  &0.46708(6)   & 0.30104(7) & -0.0360(1) & 0.0179(1) \\ 
    1  &0.3593(1)   & 0.32367(5) & -0.01224(3) & -0.0027(1) \\ 
    2  &0.34310(5)   & 0.33548(4) & -0.01065(5) & -0.0089(1) \\ 
    3  &0.3411(2)  & 0.33906(8) & -0.0109(1) & -0.0106(1) \\ 
    4  &0.3411(3)  & 0.3401(1) &  -0.0111(2) & -0.0112(2) \\ 
    5  &0.3411(2)  & 0.3402(1) &  -0.0112(1) & -0.0114(2) \\ 
    \hline
    \hline
  \end{tabular}
  \caption{$\frac{P}{\beta}=2.0031299$}
\end{subtable}%
\\
\begin{subtable}[t]{0.48\textwidth}
  \begin{tabular}{lllll} 
    \hline
    \hline
    $n$ & $K^{(n)}_{1,x}$ & $K^{(n)}_{1,y}$ & $K^{(n)}_{4,x}$ & $K^{(n)}_{4,y}$\\
    \hline
    0  &0.46681(5)   & 0.30126(8) & -0.0362(1) & 0.0180(1) \\ 
    1  &0.3590(1)   & 0.3242(1) & -0.0123(1) & -0.0027(1) \\ 
    2  &0.3430(1)   & 0.33586(6) & -0.01074(5) & -0.0088(1) \\ 
    3  &0.3407(2)  & 0.3394(3) & -0.0110(1) & -0.0106(1) \\ 
    4  &0.3408(3)  & 0.3406(2) & -0.0112(2) & -0.0110(2) \\ 
    5  &0.3410(3)  & 0.3408(2) & -0.0112(1) & -0.0112(1) \\ 
    \hline
    \hline
  \end{tabular}
  \caption{$\frac{P}{\beta}=2$}
\end{subtable}%
\\
\begin{subtable}[t]{0.48\textwidth}
  \begin{tabular}{lllll} 
    \hline
    \hline
    $n$ & $K_{1,x}$ & $K_{1,y}$ & $K_{4,x}$ & $K_{4,y}$\\
    \hline
    0  &0.46579(6)  & 0.30175(7) & -0.0361(1)  & 0.0179(1) \\ 
    1  &0.3584(1)   & 0.3245(1)  & -0.0124(1)  & -0.0026(1) \\ 
    2  &0.3423(1)   & 0.3366(1)  & -0.0107(1)  & -0.0088(1) \\ 
    3  &0.3405(2)   & 0.3398(2)  & -0.0110(1)  & -0.0105(1) \\ 
    4  &0.3404(2)   & 0.3409(2)  & -0.0113(2)  & -0.0110(2) \\ 
    5  &0.3402(3)   & 0.3409(2)  & -0.0113(2)  & -0.0111(2) \\ 
    \hline
    \hline
  \end{tabular}
  \caption{$\frac{P}{\beta}=1.99688$}
\end{subtable}
\caption{The renormalized constants for the $d=1$ TFIM.
For each $n$, $L = P = 8 \times 2^n$.
VMCRG is done with 4000 variational steps. 
During each variaional step, the MC sampling is done on 16 cores in parallel, where each core does MC sampling of 20000 Wolff steps. 
The optimization step is $\mu = 0.001$. 
The number in the paranthesis is the uncertainty on the last digit.}
\label{table:tfim_1d}
\end{table}

For $d=2$ TFIM model, we look for the value of $P/\beta$ where switching of the renormalized constants in space and time occurs at large $n$. 
The comparison is only done for the nearest neighbor coupling, which has the smallest statistical uncertainty.  
Thus, we present in the tables only the nearest neighbor coupling constants that correspond to the last RG iteration. 
The $K^{(n)}$s calculated by VMCRG for the $d=2$ TFIM, simulated at $g = g_c = 3.04438$ \cite{Continuous_time_II}, are reported in Table \ref{table:tfim_2d}. 
They lead to an estimate of the sound velocity in the interval (3.40, 3.42).
\begin{table}
  \setlength{\tabcolsep}{2.2em}
  \begin{tabular}{lll} 
    \hline
    \hline
    $\frac{P}{\beta}$ & $K^{(4)}_{1,x}$ & $K^{(4)}_{1,yz}$\\
    \hline
    3.42246  &0.1603(3) & 0.1597(2) \\ 
    3.40426  &0.1594(3) & 0.1602(2) \\ 
    \hline
    \hline
  \end{tabular}
\caption{The renormalized constants for the $d=2$ TFIM. 
$L = P = 128$.  
$K^{(4)}_{1,x}$ and $K^{(4)}_{1,yz}$ are respectively the renormalized nearest neighbor spin constants along the time and the space direction at $n = 4$.  
}
\label{table:tfim_2d}
\end{table}

\subsection{$Q=3$ and $4$}
When $d = 1$, and $Q = 3$ or $4$, the Potts model experiences a continuous phase transition at $g_c = 1$, exhibiting conformal invariance \cite{Baxter_Potts}.  
A sound velocity is thus well-defined at criticality. 
The spin variable is $\sigma = 0$ or 1, and we use coupling terms listed in Table \ref{table:potts_S}. 
We report the calculated renormalized constants $K^{(n)}$s in Table \ref{table:potts_K_q3} and \ref{table:potts_K_q4}.   
The sound velocity is determined with the nearest neighbor coupling at the last RG iteration. 
\begin{table}
  \setlength{\tabcolsep}{0.2em}
\centering
  \begin{tabular}{ll} 
    \hline
    \hline
    $\alpha$\hspace{5mm} & Coupling term\\
    \hline
    1$,x$  & $\delta_{\sigma_i\sigma_j}$ for 1st neighobor $i, j$ along the time direction \\ 
    1$,y$  & $\delta_{\sigma_i\sigma_j}$ for 1st neighobor $i, j$ along the  space direction \\ 
    2      & $\delta_{\sigma_i\sigma_j}$ for 2nd neighbor $i, j$\\ 
    3$,x$  & $\delta_{\sigma_i\sigma_j}$ for 3rd neighobor $i, j$ along the time direction \\ 
    3$,y$  & $\delta_{\sigma_i\sigma_j}$ for 3rd neighobor $i, j$ along the space direction \\ 
    \hline
    \hline
  \end{tabular}
  \caption{The couplings used for $d = 1$, $Q=3$ and $4$ Potts model.
  Note that when $\alpha = 2$, the coupling is itself isotropic between space and time. 
  }
\label{table:potts_S}
\end{table}
\begin{table}
  \setlength{\tabcolsep}{0.8em}
\centering
\begin{subtable}[t]{0.48\textwidth}
  \begin{tabular}{lllll} 
    \hline
    \hline
    $n$ & $K^{(n)}_{1,x}$ & $K^{(n)}_{1,y}$ & $K^{(n)}_{3,x}$ & $K^{(n)}_{3,y}$\\
    \hline
    0  &1.068(2)    & 0.6701(4)  & -0.0651(6)   & 0.0343(9) \\ 
    1  &0.8158(4)   & 0.7323(5)  & -0.0267(1)   & -0.0076(4) \\ 
    2  &0.766(1)    & 0.749(2)   & -0.0260(8)   & -0.022(1) \\ 
    3  &0.754(1)    & 0.750(1)   & -0.025(1)    & -0.025(1) \\ 
    4  &0.7477(4)   & 0.7468(4)  & -0.0255(6)   & -0.0251(6)\\ 
    5  &0.7452(2)   & 0.7446(2)  & -0.0252(4)   & -0.0250(4)\\ 
    \hline
    \hline
  \end{tabular}
  \caption{$\frac{P}{\beta}=2.5995$}
\end{subtable}%
\\
\begin{subtable}[t]{0.48\textwidth}
  \begin{tabular}{lllll} 
    \hline
    \hline
    $n$ & $K^{(n)}_{1,x}$ & $K^{(n)}_{1,y}$ & $K^{(n)}_{3,x}$ & $K^{(n)}_{3,y}$\\
    \hline
    0  &1.067(2)   & 0.6714(5) & -0.0653(5) & 0.0350(8) \\ 
    1  &0.8150(4)  & 0.7350(5) & -0.0278(2) & -0.0079(5) \\ 
    2  &0.765(1)   & 0.751(2)  & -0.0252(7) & -0.022(1) \\ 
    3  &0.752(1)   & 0.750(1)  & -0.025(1) & -0.024(1)\\ 
    4  &0.7469(4)  & 0.7479(5) & -0.0250(6) & -0.0248(3)\\ 
    5  &0.7441(3)  & 0.7456(3) & -0.0253(4) & -0.0251(2)\\ 
    \hline
    \hline
  \end{tabular}
  \caption{$\frac{P}{\beta}=2.5942$}
\end{subtable}%
\caption{The renormalized constants for the $d=1$, $Q=3$ Potts model.
For each $n$, $L = P = 8 \times 2^n$.
When $n = 0$ to $3$, the simulations are done with Metropolis local updates with 1000 variational steps. 
When $n = 0, 1, 2$, each variational step uses 100 sweeps of MC averaging in parallel on 8 cores.  
When $n = 3$, each variational step uses 500 sweeps. 
For $n = 4$ and 5, the simulation details are the same as in Table \ref{table:tfim_1d}. 
}
\label{table:potts_K_q3}
\end{table}

\begin{table}
  \setlength{\tabcolsep}{0.8em}
\centering
\begin{subtable}[t]{0.48\textwidth}
  \begin{tabular}{lllll} 
    \hline
    \hline
    $n$ & $K^{(n)}_{1,x}$ & $K^{(n)}_{1,y}$ & $K^{(n)}_{3,x}$ & $K^{(n)}_{3,y}$\\
    \hline
    0  & 1.171(2)  & 0.7188(5) & -0.0615(4) & 0.033(1) \\ 
    1  & 0.872(2)  & 0.777(1)  & -0.027(1)  & -0.0084(4)\\ 
    2  & 0.801(1)  & 0.781(2)  & -0.024(1)  & -0.021(1) \\ 
    3  & 0.770(1)  & 0.765(1)  & -0.023(1)  & -0.023(1)\\ 
    4  & 0.7519(5) & 0.7498(4) & -0.0225(2) & -0.0225(2)\\ 
    5  & 0.7374(3) & 0.7355(5) & -0.0217(3) & -0.0215(3) \\ 
    \hline
    \hline
  \end{tabular}
  \caption{$\frac{P}{\beta}=3.146$}
\end{subtable}%
\\
\begin{subtable}[t]{0.48\textwidth}
  \begin{tabular}{lllll} 
    \hline
    \hline
    $n$ & $K^{(n)}_{1,x}$ & $K^{(n)}_{1,y}$ & $K^{(n)}_{3,x}$ & $K^{(n)}_{3,y}$\\
    \hline
    0  & 1.167(2)  & 0.7206(4) & -0.0612(4) & 0.033(2)\\ 
    1  & 0.872(2)  & 0.779(1)  & -0.0255(9) & -0.0081(4)\\ 
    2  & 0.800(1)  & 0.782(2)  & -0.022(1)  & -0.019(1) \\ 
    3  & 0.769(1)  & 0.765(1)  & -0.022(1)  & -0.024(1) \\ 
    4  & 0.7500(4) & 0.7508(3) & -0.0226(2) & -0.0221(2)\\ 
    5  & 0.7355(3) & 0.7372(3) & -0.0217(4) & -0.0216(4)\\ 
    \hline
    \hline
  \end{tabular}
  \caption{$\frac{P}{\beta}=3.137$}
\end{subtable}%
\caption{The renormalized constants for the $d=1$, $Q=4$ Potts model.
For each $n$, $L = P = 8 \times 2^n$.
The simulation details are the same as in Table \ref{table:potts_K_q3}. 
}
\label{table:potts_K_q4}
\end{table}

This estimates $v_s$ in the interval $(2.594, 2.600)$ for $Q=3$, and $(3.137, 3.146)$ for $Q=4$, to be compared with the analytical result $v_s = \pi$ when $Q = 4$ \cite{Lajko_EE}. 
For comparison, fitting the finite size behavior of the critical free energy against the CFT prediction \cite{Affleck_CFT, Blote_CFT} leads to $v_s$ = 2.598 for $Q =3$ and $v_s$ = 3.156 for $Q = 4$ \cite{Gehlen_CFT}. 
Fitting the finite size behavior of the critical entanglement entropy against the CFT prediction \cite{EE_QFT} leads to $v_s = 2.513$ for $Q = 3$ and $v_s = 2.765$ for $Q = 4$ \cite{Ma_CFT}.

We observe that the (approximate) space-time isotropy occurs before the fixed-point Hamiltonian is reached.   
For example, in the $Q = 4$ Potts model, it is known that a logarithmic scaling operator is present around the fixed-point Hamiltonian, which makes the approach to the fixed-point Hamiltonian very slow. 
This is indeed what one sees in Table \ref{table:potts_K_q4}.  
However, as along as this scaling operator is isotropic, one expects that the slow approach to the fixed-point Hamiltonian should not affect the convergence to isotropy. 
This is also what one sees. 
The sound velocity can therefore be obtained with less RG iterations than requried for computing, say, the critical exponents of the model. 

\section{The energy-stress tensor}
\label{sec:stress}
As one changes the parameter $\frac{P}{\beta}$ in the zeroth RG iteration, one also changes the fixed-point Hamiltonian reached by the RG procedure, as shown, for example, in Table \ref{table:tfim_1d_K}.   
Since dilational transformations are isotropic, there is a line of fixed-point Hamiltonians reflecting the different extent of anisotropy in the system \cite{Cardy}. 
A change of $\frac{P}{\beta}$ generates a movement along this line of fixed-point Hamiltonians. 
In fact, fixing $P$, the change $\beta \rightarrow  \beta + \delta \beta$ induces a coordinate transformation: $x_0 \rightarrow x'_0  = (1 - \frac{\delta\beta}{\beta})x_0, x_1 \rightarrow x'_1 = x_1$, where $x_0$ and $x_1$ are time and space coordinates, respectively. 
Here we have taken the coordinate transformation to be passive, i.e. $\vec x = (x_0, x_1)$ and $\vec x' = (x'_0, x'_1)$ denote the number of lattice spacings needed to describe the same physical length, before and after the transformation.  
Thus, a time dilation generates a change in the system Hamiltonian. 
In field theory, the response of the system Hamiltonian to a generic coordinate transformation, $x^\mu \rightarrow x'^{\mu} = x^\mu + \epsilon^\mu(\vec x)$, is described by the energy-stress tensor, $T^{\mu\nu}$, defined by  
\begin{equation}
  \delta H = - \frac{1}{(2\pi)^{D-1}} \int \frac{\partial \epsilon^\mu}{\partial x_\nu} T_{\mu\nu} d^D x
  \label{eq:T}
\end{equation}
where $D$ is the space-time dimension of the system. 
As $\beta$ is conjugate to $\hat H$ in the action, we identify $T_{00}$ as the energy operator in the path integral. 

To appreciate the novelty brought by VMCRG in this context, let us consider, for example, the two-dimensional classical Ising model with the Hamiltonian
\begin{equation}
  H_{\Ising}(\bm\sigma) = -K_{0} \sum_{\braket{i,j}_{0}} \sigma_i\sigma_j - K_{1} \sum_{\braket{i,j}_{1}} \sigma_i\sigma_j
\end{equation}
where $\braket{i,j}_{0}$ and $\braket{i,j}_{1}$ are nearest neighbor spins along the $x_0$ and the $x_1$ direction, respectively. 
The system is isotropic and critical when $K_0 = K_1 = K_c = \text{arcsinh}(1)/2 = 0.4407\cdots$ \cite{Schultz_Ising}.  
An infinitesimal change in the coupling constant, $K_0 = K_c - \delta J$ and $K_1 = K_c + \delta J$, turns on anisotropy yet the system still maintains its criticality.    
That is, the deviation from the isotropic Hamiltonian,  
\begin{equation}
  \begin{split}
    \delta H(\bm\sigma) &= \sum_{m,n} A_{m,n}(\bm\sigma)\delta J 
  \\
  &\equiv  \sum_{m,n} (\sigma_{m,n} \sigma_{m+1,n} - \sigma_{m,n} \sigma_{m, n+1}) \delta J
\end{split}
  \label{eq:A}
\end{equation}
generates a length scale transformation $x_0 \rightarrow x'_0 = (1 - \delta\lambda) x_0$ and $x_1 \rightarrow x'_1 = (1+\delta \lambda)x_1$, where $\delta \lambda = \frac{1}{\gamma} \delta J$ with an unknown proportionality constant $\gamma$. 
Continuous-time VMCRG provides a way to determine $\gamma$ directly, and in that sense, it is a ruler of anisotropy. 

To determine $\gamma$, we invoke the universality of the fixed-point Hamiltonians. 
Let $H^* (\bm\mu)$ be the fixed-point Hamiltonian that VMCRG eventually reaches, starting from the critical $d=1$ TFIM with $\frac{P}{\beta} = v_s$ and from the critical isotropic $d=2$ classical Ising model.  
In practice, we approximate $H^*(\bm\mu)$ with $H^{(n)}(\bm\mu)$ for some large $n$. 
For the TFIM, the change in the action $\beta \hat H \rightarrow (\beta + \delta\beta) \hat H$ generates a change in the fixed-point Hamiltonian $\delta H^{(n)}(\bm\mu) = - \sum_{\alpha} \frac{\partial K^{(n)}_\alpha}{\partial \beta }S_\alpha(\bm\mu)\delta \beta$ with an anisotropy of the extent $\delta(\frac{x_1}{x_0}) = \frac{x'_1}{x'_0} - \frac{x_1}{x_0} = \frac{\delta \beta}{\beta} \frac{x_1}{x_0}$. 
For the classical $d=2$ Ising model, the change in the unrenormalized Hamiltonian $H_\Ising \rightarrow H_\Ising + \sum_{m,n}A_{m,n}\delta J$ generates a change in the fixed-point Hamiltonian $\delta H^{(n)}(\bm\mu) = -\sum_\alpha \frac{\partial K^{(n)}_\alpha}{\partial J}S_\alpha(\bm\mu)\delta J$, with an anisotropy of the extent $\delta(\frac{x_1}{x_0}) = 2\delta \lambda \frac{x_1}{x_0}$.  
The $\delta H^{(n)}(\bm\mu)$ for the TFIM and for the $d=2$ classical Ising model should be multiples of each other, because the line of fixed-point Hamiltonians is universal.   
In particular, when they are equal, the anisotropies that they represent should coincide. 
This means that 
\begin{equation}
\gamma = \frac{\delta J}{\delta\lambda} = 2\,\frac{\beta \,\partial K_\alpha^{(n)}/\partial \beta}{\partial K_\alpha^{(n)}/\partial J}
\label{eq:gamma}
\end{equation}
for all $\alpha$. 
The Jacobians of the RG transformation, $\frac{\partial K_\alpha^{(n)}}{\partial J}$ and $\frac{\partial K_\alpha^{(n)}}{\partial \beta}$, can be readily computed with VMCRG \cite{VMCRG}, where the first Jacobian is calculated for the TFIM, and the second one for the classical Ising model.
For the operator $A(\bm\sigma)$ defined in Eq. \ref{eq:A}, $\gamma$ is analytically known and is $\frac{\sqrt{2}}{2} = 0.7071\cdots$.  
A VMCRG calculation using Eq. \ref{eq:gamma} with $n = 4$ gives $\gamma = 0.708 \pm 0.001$, so the ruler works. 

With the coordinate transformation $x_0 \rightarrow x'_0 = (1 - \delta\lambda) x_0$ and $x_1 \rightarrow x'_1 = (1+\delta \lambda)x_1$, a part of the energy-stress tensor can now be read off from Eq. \ref{eq:T}: 
\begin{equation}
  \sum_{m,n} A_{m,n}(\bm\sigma) \delta J  = \delta H(\bm\sigma) = - \frac{1}{2\pi} \int (-\delta \lambda T_{00} + \delta \lambda T_{11}) d^2x.  
\end{equation}
We take the lattice spacing to be 1, and $\sum_{m,n}$ is equivalent to $\int d^2x$. 
This gives
\begin{equation}
  \gamma A = \frac{1}{\pi} (T + \bar{T}), 
  \label{eq:gA}
\end{equation}
where, in 2D, $T$ and $\bar{T}$ are respectively the holomorphic and the antiholomorphic component of the energy-stress tensor, and are defined as $T = \frac{1}{4}(T_{00} - 2 i T_{01} - T_{11})$ and $\bar{T} = \frac{1}{4}(T_{00} + 2 i T_{01} - T_{11})$.   
While the argument is developed for the Ising model, it also generalizes to other 2D systems. 

A consequence of Eq. \ref{eq:gA} is that one obtains a prediction of the finite-size dependence of $\braket{A}$, due to CFT.  
For example, if one simulates a critical system infinitely long along the $x_0$ direction but periodic of size $L$ along $x_1$, CFT predicts that $\braket{T} = \braket{\bar{T}} = -(\frac{2\pi}{L})^2\frac{c}{24}$, and thus $\braket{A} = -\frac{1}{\gamma \pi} (\frac{2\pi}{L})^2 \frac{c}{12}$, where $c$ is the central charge of the underlying CFT.  
This prediction on $A$ has been verified in \cite{CFT_Stress}.  

\section{Conclusion}
\label{sec:conclude}
In this paper, we have shown how to perform MCRG with continuous-time Monte Carlo simulations, and demonstrated that space-time isotropy is explicitly recovered at large distances.  
This yields a practical method to determine the sound velocity and the energy-stress tensor from their defining expressions. 
This should allow generalizations to systems in three dimensions, which could be studied in the future. 
\begin{acknowledgments}
The authors acknowledge support from the DOE Award DE-SC0017865. 
\end{acknowledgments}
\bibliographystyle{apsrev}
\bibliography{abc}
\end{document}